\documentclass[prl,singlecolumn, nopacs, preprint]{revtex4}
\usepackage{graphicx}
\usepackage{graphics}
\usepackage{amsfonts}
\usepackage{amsmath}
\usepackage{epsfig}
\usepackage{color}

\begin{document}
\catcode`\ä = \active \catcode`\ö = \active \catcode`\ü = \active
\catcode`\Ä = \active \catcode`\Ö = \active \catcode`\Ü = \active
\catcode`\ß = \active \catcode`\é = \active \catcode`\è = \active
\catcode`\ë = \active \catcode`\ô = \active \catcode`\ê = \active
\catcode`\ø = \active \catcode`\ò = \active \catcode`\í = \active
\catcode`\Ó = \active \catcode`\ú = \active \catcode`\á = \active
\catcode`\ã = \active
\defä{\"a} \defö{\"o} \defü{\"u} \defÄ{\"A} \defÖ{\"O} \defÜ{\"U} \defß{\ss} \defé{\'{e}}
\defè{\`{e}} \defë{\"{e}} \defô{\^{o}} \defê{\^{e}} \defø{\o} \defò{\`{o}} \defí{\'{i}}
\defÓ{\'{O}} \defú{\'{u}} \defá{\'{a}} \defã{\~{a}}



\newcommand{\li}{$^6$Li}
\newcommand{\na}{$^{23}$Na}
\newcommand{\cs}{$^{133}$Cs}
\newcommand{\kk}{$^{40}$K}
\newcommand{\rb}{$^{87}$Rb}
\newcommand{\vect}[1]{\mathbf #1}
\newcommand{\g}{g^{(2)}}
\newcommand{\one}{$\left|1\right>$}
\newcommand{\two}{$\left|2\right>$}
\newcommand{\V}{V_{12}}
\newcommand{\kfa}{\frac{1}{k_F a}}

\title{Fermionic Superfluidity with Imbalanced Spin Populations \\and the\\ Quantum Phase Transition to the Normal State}

\author{Martin W. Zwierlein}\email[to whom correspondence should be addressed. E-mail:]{zwierlei@mit.edu}
\author{Andr\'e Schirotzek}
\author{Christian H. Schunck}
\author{Wolfgang Ketterle}

\affiliation{Department of Physics\mbox{,} MIT-Harvard Center for
Ultracold Atoms\mbox{,} and Research Laboratory of Electronics,\\
MIT, Cambridge, MA 02139}

\date{\today}

\def\abstractname{}
\begin{abstract}
\vspace{20mm}
\begin{bf}
Whether it occurs in superconductors, helium-3 or inside a neutron
star, fermionic superfluidity requires pairing of fermions,
particles with half-integer spin. For an equal mixture of two
states of fermions ("spin up" and "spin down"), pairing can be
complete and the entire system will become superfluid. When the
two populations of fermions are unequal, not every particle can
find a partner. Will the system nevertheless stay superfluid? Here
we study this intriguing question in an unequal mixture of
strongly interacting ultracold fermionic atoms. The superfluid
region vs population imbalance is mapped out by employing two
complementary indicators: The presence or absence of vortices in a
rotating mixture, as well as the fraction of condensed fermion
pairs in the gas. Due to the strong interactions near a Feshbach
resonance, the superfluid state is remarkably stable in response
to population imbalance. The final breakdown of superfluidity
marks a new quantum phase transition, the Pauli limit of
superfluidity.

\end{bf}
\end{abstract}

\maketitle

The study of superfluidity in an unequal mixture of two Fermi
gases, or more generally unequal Fermi surfaces, is highly
relevant to a wide area of physics. For example, this situation
should occur in strongly degenerate quark matter in the core of a
neutron star, where quarks of differing mass will belong to Fermi
spheres of differing size. The ground state of such a system has
been the subject of debate for
decades~\cite{clog62,FF64,LO64,sarm63,liu03} and experimental
studies are highly desirable. However, the experimental realization
of unbalanced Fermi seas in superconductors, charged fermionic
superfluids, poses extreme difficulties: To reach imbalanced
electron densities of spin up vs spin down electrons, a natural
idea would be to apply a magnetic field. However, magnetic fields
are either fully shielded from the superconductor via the Meissner
effect, or they enter only in the form of quantized flux lines or
vortices. Experiments trying to study mismatched Fermi surfaces
therefore have to suppress these effects, as in
experiments in heavy fermion
superconductors~\cite{casa04,rado03,bian03} or
quasi-two-dimensional organic superconductors~\cite{casa04}.
In the neutral superfluid helium-3, one can mismatch the Fermi
surfaces by a magnetic field and thus destroy inter-spin pairing.
However, superfluidity persists due to (p-wave) pairing between
equal spins~\cite{voll90}.

The recently discovered atomic fermionic
superfluids~\cite{rega03mol,joch03bec,zwie03molBEC,bart04,rega04,zwie04rescond,bour04coll,kina04sfluid,bart04coll,chin04gap,kina05heat,part05,zwie05vort}
provide an exciting new possibility to explore unequal mixtures of
fermions. Here, populations in two hyperfine states of the
fermionic atom can be freely chosen. In
addition, the (s-wave) interactions between two atoms in
different states and hence the binding energy of atom pairs can be
tuned. In equal mixtures of fermions, this tunability is
being exploited to study the crossover from a Bose-Einstein
Condensate (BEC) of molecules to a Bardeen-Cooper-Schrieffer (BCS)
superfluid of loosely bound
pairs~\cite{bart04,bour04coll,rega04,zwie04rescond,chin04gap,part05,zwie05vort}.
At zero temperature, this crossover is smooth
~\cite{eagl69,legg80,nozi85}, the system stays superfluid even
for arbitrarily weak interaction and no phase transition occurs.
In the case of unequal mixtures, the phase diagram is predicted to
be much richer~\cite{carl05,sedr05,pao05,shee05,son05,yang05}. In the
molecular limit of tight binding, all fermions in the less
populated spin state will pair up with atoms in the other state. The resulting molecular condensate will spatially
coexist with the remaining Fermi sea of unpaired atoms. As
the repulsive interaction between atoms and molecules is increased, the condensate will start to expel unpaired atoms,
leading to a phase separation of the superfluid from the normal
phase~\cite{vive00,powe05bosefermi,pao05,carl05,shee05}. This picture is
expected~\cite{beda03,pao05,carl05,shee05} to extend into the
BCS-limit of weakly bound pairs. Here, the pairing gap $\Delta$
prevents unpaired atoms from entering the BCS superfluid~\cite{beda03}.
As the binding and hence the pairing gap is further reduced,
$\Delta$ will eventually become small compared to the chemical
potential difference $\delta \mu = \mu_2 - \mu_1$ between the two
spin states, allowing unpaired excess atoms to enter the
superfluid region. Close to this point, superfluidity will cease
to exist~\cite{footnote:FFLO}. In the weakly
interacting BCS-limit the pairing gap is exponentially small
compared to the Fermi energy, hence an exponentially small
population imbalance can destroy superfluidity.

This superfluid to normal transition is an example of a quantum
phase transition, which occurs even at zero temperature, when
all thermal fluctuations are frozen out and only quantum
fluctuations prevail. It can also be driven by increasing the
mismatch in chemical potentials between the two spin states to the
critical value of $\delta\mu \approx \Delta$, inducing collapse into the normal
state. In this context the phase transition is known as the Pauli
or Clogston limit of superfluidity~\cite{clog62}. Its exact nature, whether there is one or several first-order and/or second-order transitions, however, is still the subject of current
debate~\cite{son05,yang05,casa04}.

In this paper we map out the superfluid region as a function of
population imbalance, interaction strength and temperature in an ultracold
fermionic gas of \li\ atoms. The only direct and unambiguous
signature of superfluid flow in Fermi gases so far is the presence
of vortices~\cite{zwie05vort}. By studying unequal Fermi mixtures
under rotation we establish superfluidity for a broad range in the population mismatch. Close to the breakdown of superfluidity, vortices are
strongly damped and difficult to observe.  Therefore, we map out
the full regime of superfluidity by determining the fraction of
condensed fermion pairs in a non-rotating
cloud~\cite{rega04,zwie04rescond}.

The two experimental methods require slightly different procedures
for imaging the pair condensate wavefunction after release from
the trap. To extract the fraction of condensed vs uncondensed
pairs, the condensate must separate well from the thermal cloud
and should therefore remain small. For the detection of rotating
clouds, the condensate should expand to a large size in order to
magnify the vortices. In the following, we give the parameters
used to determine the condensate fraction in parentheses
after those used for vortex detection.

In the experiment, fermionic \li\ atoms were sympathetically
cooled to degeneracy by \na\ atoms in a magnetic
trap~\cite{hadz03big_fermi}. The ultracold cloud was subsequently
loaded into an optical dipole trap (waist $w \approx 120\, \mu $m) at a
maximum trap depth of about 8 $\mu$K. At a magnetic bias field of
875 G, a variable spin-mixture of the two lowest hyperfine states
(labelled \one\ and \two) was created via a Landau-Zener sweep
with variable sweep rate. Interactions between these two states are
strongly enhanced around a 300 G wide Feshbach resonance located
at 834 G~\cite{bart04fesh}. At lower values of the magnetic bias field, two isolated
fermions can bind into a stable molecule, while at higher values
fermion pairs can only exist in the stabilizing presence of the surrounding gas. This tunability of the
binding energy provides access to the BEC-BCS crossover physics.
The spin mixture was evaporatively cooled further by lowering the
trap depth to 1.6 $\mu$K resulting in radial and axial trap
frequencies of $\nu_r = 110$ Hz and $\nu_a = 23$ Hz, respectively.
At the same time, the magnetic field was ramped to 812 G (818 G),
which is on the BEC-side of the resonance, but still in the regime
of strong interaction. Here, $1/k_F a = 0.19\, (0.11)$, where $a$ is the
scattering length and $k_F$ is defined as the Fermi momentum of a non-interacting, equal spin mixture. The rather moderate evaporation still leaves room for thermal molecules in an equal mixture, but was chosen to ensure efficient cooling of highly asymmetric mixtures, avoiding spilling of large Fermi clouds. It ensured that the total number of atoms $N = 7 \cdot 10^6$ ($N = 2.3\, 10^7$) was approximately constant and independent of the asymmetry between the two spin states.

For the vortex experiment, we set the spin mixture in rotation
using two blue-detuned laser beams (wavelength 532 nm) rotated
symmetrically around the cloud at angular frequency $\Omega = 2\pi\,
70$ Hz)~\cite{zwie05vort}. After 800 ms of stirring, the rotating
cloud was left to equilibrate for several hundred ms.

Starting with either the rotating or the non-rotating cloud, we
then varied the interaction strength between the two spin states
in the gas by ramping the magnetic field in 100 ms (500 ms) to
several values around the Feshbach resonance (for the condensate
fraction experiment, the trap depth was simultaneously increased
to 4 $\mu$K ($\nu_r = 192$ Hz)). After 50 ms (100 ms) of hold
time, an image of the cloud was taken following the procedure
outlined in~\cite{zwie05vort}. In short, after releasing the cloud
from the optical trap the binding energy of fermion pairs was rapidly
increased by ramping the magnetic field within 2 ms (200 $\mu$s)
to 690 G, in the far wings of the resonance on the BEC side. Here, fermion
pairs were stable throughout further expansion. After a total of
11 ms (14 ms) of expansion (in the remaining magnetic saddle-point potential)
an image of either state \one\ or state \two\ was taken.
For the condensate fraction data, the magnetic field was suddenly switched to
800 G right before imaging. At this field the molecules absorb the probe light with the same
strength as free atoms.
The images revealed the center-of-mass
wavefunction of the pairs and, for rotating clouds, the eventual
presence of vortices.  For the condensate fraction experiment, the
200 $\mu$s fast ramp to the BEC-side immediately after release
from the trap ensured that even large condensates separated well from the normal, uncondensed
component. Since the ramp was fast compared to the radial trapping
period, the size of the expanded condensate was mostly governed by
the residual mean-field interaction at 690 G, where $a = 1400\,
a_0$.

\begin{figure*}
    \begin{center}
    \includegraphics[width=6.5in]{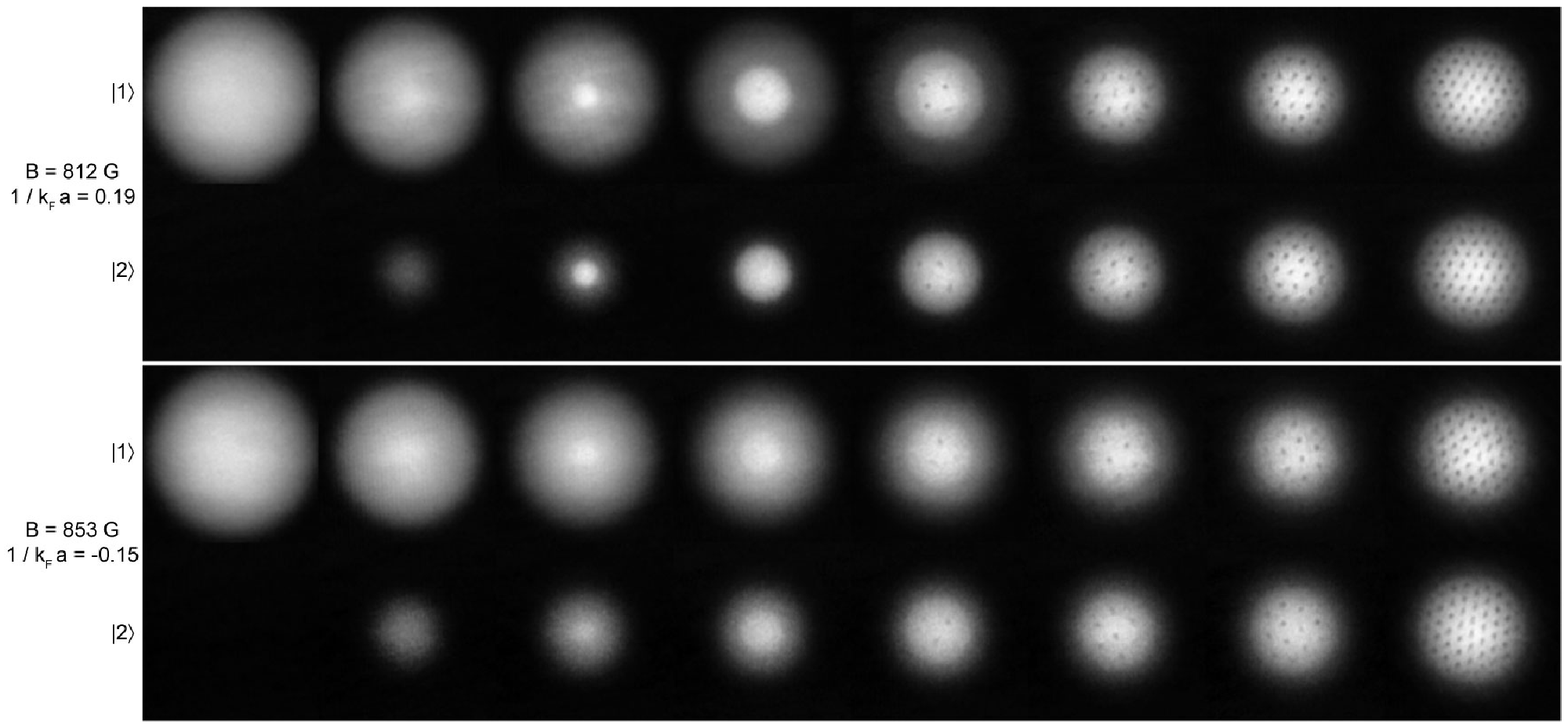}
    \caption[Title]{Superfluidity in a strongly interacting Fermi
    gas with imbalanced populations. The upper (lower) pair of
    rows shows clouds prepared at 812 G (853 G), where $1/k_F a =
    0.2$ ($1/k_F a = -0.15$). In each pair of rows, the upper
    picture shows state \one, the lower one state \two. For the
    812 G data, the population imbalance $\delta = \frac{N_2 - N_1}{N_1
    + N_2}$ between $N_1$ atoms in state \one\ and $N_2$ in state
    \two\ was (from left to right) 100\%, 90\%, 80\%, 62\%, 28\%,
    18\%, 10\% and 0\%. For the 853 G data, the mismatch was
    100\%, 74\%, 58\%, 48\%, 32\%, 16\%, 7\% and 0\%. For
    different $\delta$, the total number of atoms varied only within
    20\% around $N = 7 \cdot 10^6$, with the exception of the
    endpoints $\delta = 100\%$ ($N = 1 \cdot 10^7$) and $\delta = 0\%$ ($N =
    1.2 \cdot 10^7$). The field of view of each image was 1.4 mm $\times$ 1.4 mm.} \label{fig:Vortices}
    \end{center}
\end{figure*}

\begin{figure}
    \begin{center}
    \includegraphics[width=4.5in]{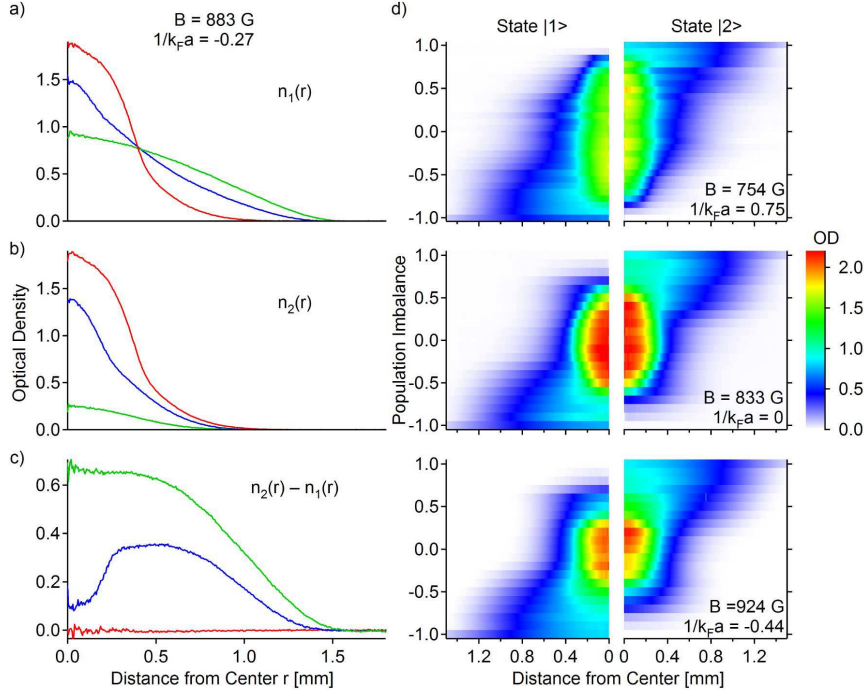}
    \caption[Title]{Radial density profiles of the two components
of a strongly interacting Fermi gas mixture with unequal
populations. a) and b) Profiles of the component in state \one\ and \two\, resp., originating from 883 G
($1/k_F a = -0.27$).  The imaging procedure is detailed in the text.
The population imbalance was $\delta = 0\%$ (red), $\delta = 46\%$ (blue) and $\delta = 86\%$
(green). c) Difference between the distributions in state \one\
and \two. The total number of atoms was $N = 2.3 \cdot 10^7$. The
clear dip in the blue curve caused by the pair condensate gives an
indication for phase separation of the superfluid from the normal
gas. d) Color-coded profiles of clouds prepared at three different
interaction strengths. The profiles are azimuthal averages of the
axially integrated density. The condensate is clearly visible as
the dense central part surrounded by unpaired fermions or
uncondensed molecules.}
    \label{fig:Cloudprofiles}
    \end{center}
\end{figure}

Fig.~\ref{fig:Vortices} shows profiles of the two spin states
for various spin-mixtures, on the BEC- and on the BCS-side of the
resonance. Starting with a pure Fermi sea in state \one\, we see how gradually, for increasing
numbers in the second spin state \two\, first a normal (uncondensed) cloud of
fermion pairs emerges, then a condensate peak appears within the normal cloud (see also Fig.~\ref{fig:Cloudprofiles}a, b). As the condensate size increases, vortices appear in the rotating cloud. Naturally, the largest
condensates with the largest number of vortices are obtained for
an equal mixture. Clearly, superfluidity in the strongly
interacting Fermi gas is not constrained to a narrow region around
the perfectly balanced spin-mixture, but instead superfluid flow
is observed for large asymmetries in the populations.

\begin{figure}
    \begin{center}
    \includegraphics[width=3.5in]{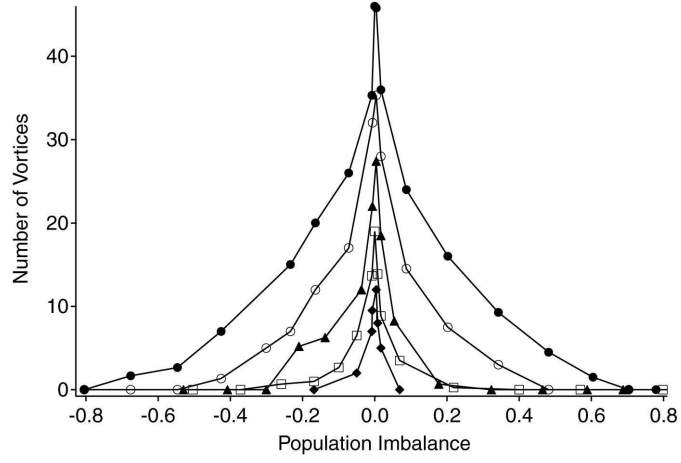}
    \caption[Title]{Vortex number vs population imbalance for different
    interaction strengths. Results are shown for 812 G or $1/k_F a =
    0.2$ (filled circles), 853 G ($1/k_F a = -0.15$, empty circles),
    874 G ($1/k_F a = -0.3$, filled triangles), 896 G ($1/k_F a =
    -0.4$, empty squares) and 917 G ($1/k_F a = -0.5$, diamonds).}
    \label{fig:nVortexvsmismatch}
    \end{center}
\end{figure}

Fig.~\ref{fig:nVortexvsmismatch} summarizes our findings for
rotating spin-mixtures. It displays the number of detected
vortices vs the population imbalance between the two spin states.
The vortex number measures qualitatively how deep the system is in
the superfluid phase: The higher the non-superfluid fraction, the
faster the condensate's rotation will damp given the non-vanishing
anisotropy ($\frac{\omega_x - \omega_y}{\omega_x+\omega_y} \approx
1.5\%$) of our trap~\cite{abos02_form,zwie05vort,guer00}. We
therefore observe how gradually, for decreasing interaction
strength on the BCS-side, the superfluid window shrinks in size around the optimal situation of equal populations.

A more detailed map of the superfluid regime as a function of
interaction strength and also temperature (see Fig.~\ref{fig:phasediagram}) was
obtained from a study of condensate fractions, which were determined
from cloud profiles such as in Fig.~\ref{fig:Cloudprofiles}.
Throughout the whole crossover region, pair condensation occurred
for a broad range of population imbalances, demonstrating again
the stability of the superfluid around the resonance.

An intriguing property of the superfluid state with imbalanced
populations is the clear depletion in the excess fermions of the
majority component, see Fig.~\ref{fig:Cloudprofiles}c. The profiles in Fig.~\ref{fig:Cloudprofiles} present
the axially integrated density, hence the true
depletion in the 3D density is even stronger. The condensate seems
to repel the excess fermions. This feature was observed after
expansion at 690 G, where interactions are still strong (initially
$1/k_F a = 2.0$). The expansion, at least in the region around the
condensate, is hydrodynamic and should proceed as a scaling
transformation~\cite{kaga97bose,meno02,cast04scal}. Therefore, the depletion observed in
ballistic expansion hints at spatial phase separation of the
superfluid from the normal state. This effect was observed
throughout the resonance region, and on resonance even when no
magnetic field ramps were performed during expansion. However, to
distinguish a phase separated state with equal densities in the
superfluid region from more exotic states allowing unequal
densities, a careful analysis of the three-dimensional density,
reconstructed from the integrated optical densities, is necessary
and will be the subject of a future study.

We did not observe (by simultaneously imaging along the long and short axis)
a modulation in the condensate density as would be
predicted for the FFLO state~\cite{mizu05fflo,sedr05,casto05}. However, this
state is predicted to be favored only in a narrow region of
parameter space and might have escaped our attention.

\begin{figure}
    \begin{center}
    \includegraphics[width=4.5in]{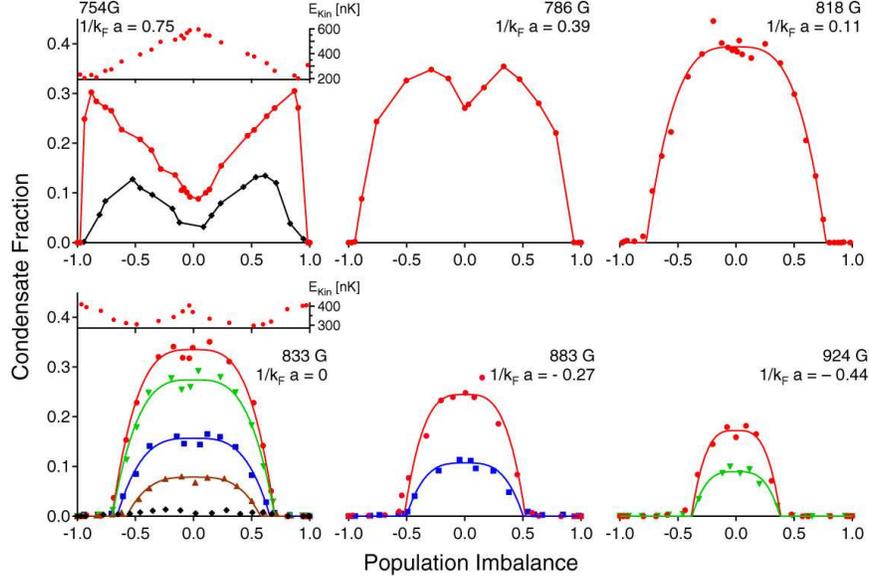}
    \caption[Title]{Condensate fraction vs population imbalance
    for several temperatures and interaction strengths. The total
    number of atoms $N = 2.3 \cdot 10^7$ is constant to within 20\% for
    all data-points ($T_F = 1.9\, \mu K$ for an equal mixture). For a given population imbalance, the
    uppermost curves for different magnetic fields are
    approximately isentropically connected. The different symbols
    correspond to different evaporation ramps. The average radial kinetic energy per molecule of thermal clouds in the minority component serves as an indicator for temperature and is shown in the insets for
    754 G and 833 G for the coldest data. On resonance, for a population asymmetry of 50\%, we measure an energy of $k_B \cdot$ 300 nK
    (circles), 345 nK (inverted triangles), 390 nK (squares), 420
    nK (triangles) and 505 nK (diamonds). The critical population
    imbalance $\delta_c$ for the breakdown of condensation at 754 G is about
    $\delta^{754}_c \approx 96\%$ and at 786 G it is $\delta^{786}_c \approx 95\%$. For the data at higher magnetic
    fields we determine $\delta_c$ through a threshold fit to the
    first three data points with non-zero condensate fraction for
    each sign of asymmetry. Although we could have used any reasonable threshold function, empirically, it was found that the function $n_c (1 - {\left|\delta/\delta_c\right|}^{3.3})$
    ($n_c$ - maximum condensate fraction) provided a good fit to
    all data points. Therefore it was used for the threshold fits and is shown as a guide to the eye.}
    \label{fig:condfracvsmismatch}
    \end{center}
\end{figure}

\begin{figure}
    \begin{center}
    \includegraphics[width=4.5in]{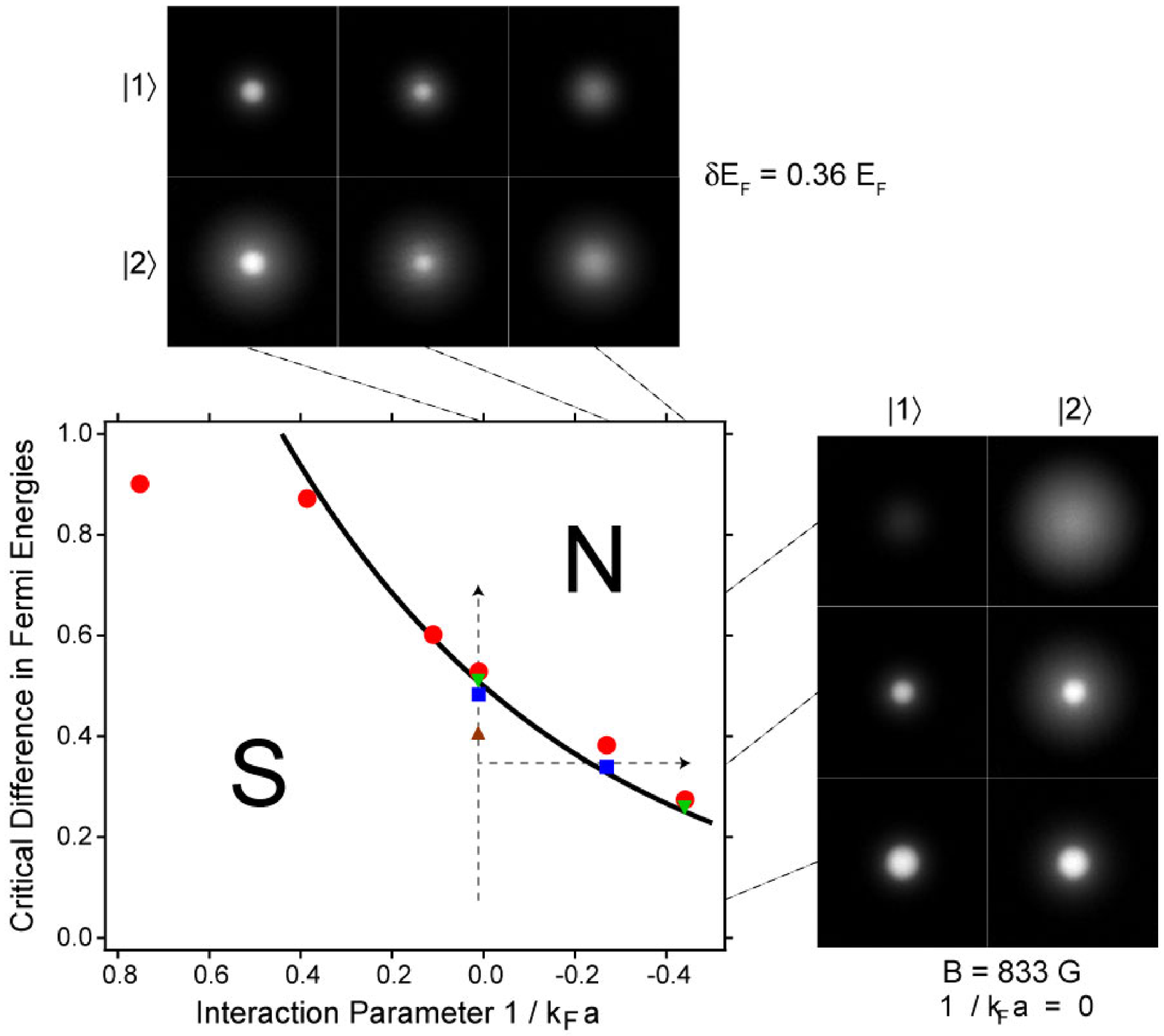}
    \caption[Title]{Critical difference in Fermi
    energies $\delta E_F$ between the two spin states for which
    the superfluid to normal transition is observed. $\delta E_F$
    for each interaction strength and temperature is obtained from
    the critical population imbalance determined in
    Fig.~\ref{fig:condfracvsmismatch} using $\delta E_F / E_F =
    (1+\delta_c)^{1/3} - (1-\delta_c)^{1/3}$. The symbols are defined in
    Fig.~\ref{fig:condfracvsmismatch}. The line shows the expected
    variation of the pairing gap $\Delta$, where the value on
    resonance has been taken from~\cite{carl05} and the
    exponential behavior in the BCS-regime, $\Delta \sim
    e^{-\pi/2k_F\left|a\right|}$ was assumed. While the trend of $\delta E_F$ is expected to follow that of $\Delta$,
    the close agreement is coincidental. Representative density
    profiles illustrate the quantum phase transition for fixed
    interaction and for fixed population imbalance along the dashed lines.}
    \label{fig:phasediagram}
    \end{center}
\end{figure}

The condensate fraction was determined from the minority
component, which in all cases is very well fit by a gaussian for thermal molecules and unpaired atoms, plus
a parabolic Thomas-Fermi-profile for the condensate. Fig.~\ref{fig:condfracvsmismatch} shows the
condensate fraction obtained for varying population difference and
temperature, and for magnetic fields or interaction strengths
around resonance. The data for 754 G, on the BEC-side of the
resonance, shows condensation over almost the entire range of
population imbalance. As the interaction strength is increased
towards resonance, the condensate fraction for equal mixtures
grows~\cite{zwie04rescond}. However, for large population asymmetries it disappears. The condensation window shrinks
further as we cross the resonance and move to the BCS-side (Fig.~\ref{fig:condfracvsmismatch}d-f).

The temperature varied with number imbalance as indicated in the insets of Fig.~\ref{fig:condfracvsmismatch}. The maximum for equal mixtures at 754 G is likely due to the energy release when more molecules were formed.
The observed critical population imbalance was only weakly dependent on temperature. This may
reflect that the pairing gap is only a weak function of
temperature, for temperatures well below the critical temperature for
superfluidity~\cite{abri75}. The critical imbalance at our coldest temperatures will thus essentially
coincide with its value at zero temperature.

On resonance, where the scattering length $a$ diverges, the system
is in the unitary regime~\cite{bert99,heis01,ho04uni,cohe05,ohar02science}, where the only remaining energy scales
of the system are the Fermi energies $E_{F,1}$ and $E_{F,2}$ of
the two spin components. The breakdown of superfluidity will occur for a certain universal ratio of these two or equivalently, in a harmonic trap, for a certain critical population imbalance.
We determine this universal number to be $\delta_c
\approx \pm 70(3)\%$ for our approximately harmonic trapping
potential. This corresponds to a Fermi energy difference $\delta E_F = E_{F,2} - E_{F,1} = \left((1+\delta_c)^{1/3} -
(1-\delta_c)^{1/3}\right) E_F = 0.53(3) E_F$, with $E_F$ the Fermi
energy of an equal mixture of non-interacting fermions.
The standard BCS state is predicted~\cite{clog62} to
break down for a critical chemical potential difference $\delta\mu
= \sqrt{2} \Delta$. On resonance, however, Monte-Carlo studies
predict~\cite{carl05} the superfluid breakdown to occur when
$\delta\mu = 2.0(1)\Delta = 1.0(1) E_F$. Only in
the weakly interacting regime do the chemical potentials equal the
Fermi energies. Quantitative agreement with the Monte-Carlo study
would require that $\delta\mu \approx 2\,\delta E_F$. This is not
unreasonable given the fact that interactions will reduce the chemical
potential of the minority component. In a preliminary analysis, we
indeed find close agreement with theory. 

Fig.~\ref{fig:phasediagram} summarizes our findings.  It shows
the critical mismatch in Fermi energies for which we observed the
breakdown of superfluidity as well as the pairing gap $\Delta$
versus the interaction parameter $1/k_F a$. Far on the BEC-side of the
resonance the superfluid is very robust with respect to population
imbalance. Here, pairing is dominantly a two-body process: The
smallest cloud of atoms in state \one\ will fully pair with
majority atoms in state \two\ and condense at sufficiently low
temperatures. On the BCS-side of the resonance, however, pairing
is purely a many-body effect and depends on the density of the two
Fermi clouds. As the density of the minority component becomes
smaller, the net energy gain from forming a pair condensate
will decrease. Even at zero temperature, this eventually leads to
the breakdown of superfluidity and the quantum phase transition to
the normal state. We have experimentally confirmed the
qualitative picture that fermionic superfluidity
breaks down when the difference in chemical potentials between the
two species becomes larger than the pairing gap.

In conclusion, we have observed superfluidity with imbalanced
spin populations. Contrary to expectations for the weakly interacting
case, superfluidity in the resonant region is extremely stable
versus population imbalance. As the asymmetry is increased, we observe the
quantum phase transition to the normal state, known as the Pauli
limit of superfluidity. Our observation opens up intriguing
possibilities for further studies on mismatched Fermi surfaces. 
One important aspect concerns the density distribution in the superfluid
regime. Standard BCS theory allows only equal spin densities,
which would entail complete phase separation of the superfluid from the
normal density. More exotic solutions~\cite{casa04} allow
superfluidity also with imbalanced densities.
Equally fascinating is the nature of the strongly correlated normal state slightly below resonance. For sufficient population imbalance we have the remarkable situation that bosonic molecules, stable even in isolation, do not condense at zero temperature, due to the presence of the Fermi sea.
\\

{\Large References and Notes}

\begin{thebibliography}{51}
\bibitem{footnote:FFLO}
Exotic superfluid states~\cite{casa04}, most
notably the FFLO-phase~\cite{FF64,LO64} of non-zero momentum
fermion pairs, are predicted to be more stable configurations than
typical BCS superfluidity, but eventually also these will be
destroyed and the gas will turn normal.
\end{thebibliography}
\end{document}